# What is the true dropped calls rate when in the test it was found to be zero?

M.V. Simkin[1] and J.Olness

*Telephia, Inc., 1 Beach St., San Francisco, CA 94133*

*Abstract*-- **We study the distributions of dropped calls rates for different wireless (cellular) carriers in different markets. Our statistics comprises over 700 different market/carrier combinations. We find that the dropped calls rates distribution is very close to lognormal. We derive an equation for the most probable dropped calls rate for particular carrier in particular market, which depends on the number of dropped calls observed, total number of calls and the parameters of the lognormal distribution. We apply this analysis to blocked and "no service" calls as well.**

*Index Terms*-- `Communication equipment testing,` `Land mobile radio cellular systems,` `Normal` `distributions, Probability, Statistics.`

In order to optimize their networks and to benchmark themselves against competitors, wireless network operators frequently test the networks using mobile test equipment. During the course of a so-called 'drive-test', wireless telephone calls are placed from many locations throughout the network. The test equipment records a large number of statistics associated with each call, including whether the call went through or was blocked and whether it terminated normally or was dropped. The results are used both to understand the end-user experience and to eliminate performance problems. The drooped calls rate is one of the most important parameters of the quality of service [1].

Analyzing the drive test data one can notice that for some market/carrier combinations the percentage of the dropped calls is strictly zero. It is obvious that it can't be the true dropped calls rate just because everything is imperfect. By looking closely at the data one sees that this occurs primarily in markets with small numbers of access counts as is summarized in the table below.

TABLE 1. FRACTION OF MARKETS WITH NO DROPPED CALLS OBSERVED.

| Markets with access count | Percent of markets with no dropped calls observed |
|---|---|
| Less than 100 | 36.8% |
| Between 100 and 1000 | 9.6% |
| More than 1000 | 0.8% |

From the above table it is clear that no dropped calls were observed because not enough calls had been made. In the present paper we show that an estimate of the true dropped calls rate can be obtained on the basis of the probability theory even in the case when no dropped calls has been observed.

To start we shall illustrate the use of the probability theory on a simple and related example. The following problem in one or other form can be found in any book of cute probability puzzles.

*Puzzle.* In certain city 10% of people are using drugs and 90% don't. A certain drug test gives correct result in 90% of cases and in 10% therefore wrong. An arbitrary person from that city was administered that drug test and it was positive. What is the probability that that person actually uses drugs?

*Solution.* A heuristic one. The evidence that the person uses drugs, which comes from the drug test (9 to 1) is equal to the evidence that the person does not use drugs, which comes from the drug use statistics (also 9 to 1) in the city. Therefore the probability is one half. A rigorous one. Let's administer the test to everybody in the city. It will be positive for 10% out of 90% non drug-users for which the test came wrong and for 90% of 10% drug-users for which the test came right. We arrive to the same answer.

How is this related to our problem? Directly. Instead of drug test we have a drive test, accuracy of which depends on the number of the access counts, and we have the statistics of the dropped calls rates for over 700 market/carrier combinations in our database.

Let's have a look at this statistics, which is shown in the Figure 1.

One who is familiar with Statistics can immediately notice that the distribution in the Figure 1 is lognormal [2]. This means that it will look normal if we shall use the logarithmic coordinates. That this is true one can see from Figures 2 and 3 (market/carrier where observed dropped rate was zero had to be excluded from this analysis). Apart from visual similarity the fact is supported by small values of skewness and kurtosis (see Table 2 in Appendix A).

From the Figure 3 one can see that the actual distribution of dropped calls rate can be approximated by lognormal distribution with high accuracy. This is convenient because lognormal distribution is completely described by two parameters: mean value $\langle \ln(x) \rangle$ ($x$ is the drop calls rate) and dispersion $s$. The mean value coincides with the maximum

---

of the bell-shaped curve of the normal distribution and $s$ is the half-width of this bell.

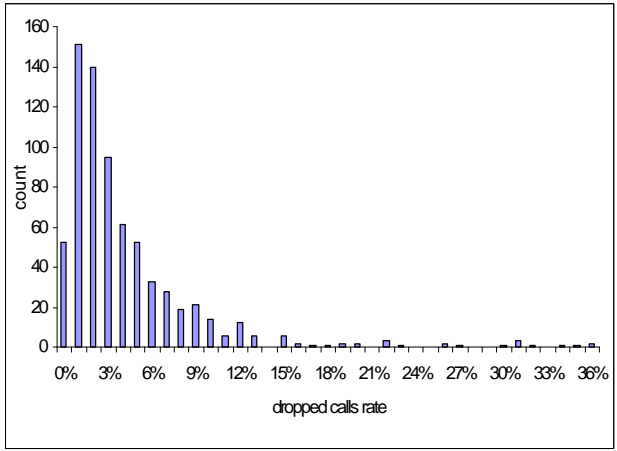

**Figure 1**. Histogram of dropped calls rates distribution for 720 market/carrier combinations.

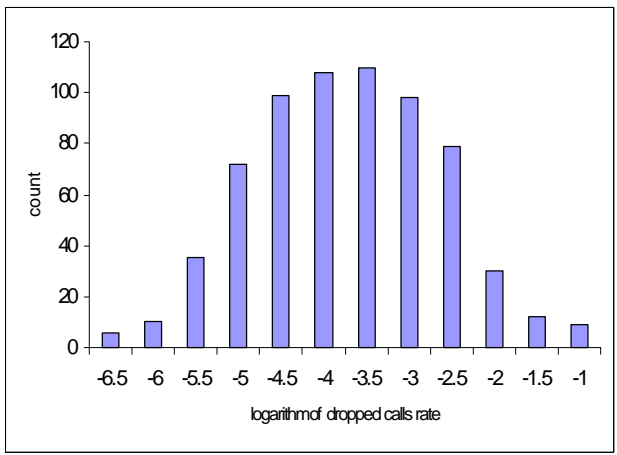

**Figure 2.** Histogram of the natural logarithm of dropped call rates distribution.

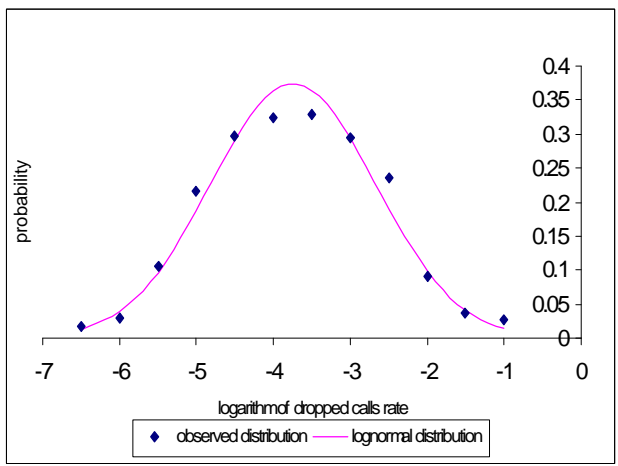

**Figure 3.** Probability density inferred from the histogram of Figure 2 compared to lognormal distribution with the same mean and standard deviation.

The formula for lognormal distribution is:

$$p(\ln(x)) = \frac{1}{\sqrt{2p}\,s} \exp(-(\ln(x) - \langle\ln(x)\rangle)^2 / 2s^2) \quad (1)$$

The values of the parameters in our case are: $\langle\ln(x)\rangle \approx -3.74$ and $s = 1.07$.

Now in analogy with the puzzle at the beginning of the paper we assume that the dropped calls rates come from this distribution, the same way as that person came from a city with known drug use statistics.

Now if we know the true dropped call rate than the probability of getting $n$ dropped calls out of $N$ calls is given by the Poisson distribution [3]:

$$p(n, x, N) = \exp(-Nx)(xN)^n / n! \quad (2)$$

Probability to observe zero dropped calls is therefore:

$$p(0, x, N) = \exp(-Nx) \quad (3)$$

Now let us consider a concrete example. For Bell South in Athens, GA with 289 access counts no dropped calls were observed. What is the true drop rate? Suppose that it is the drop rate where our distribution has a maximum at $\ln(x) \approx -3.74$ which corresponds to $x \approx 0.024$. The probability to have such drop rate is high, about 0.4. However if the drop rate is that, than the probability to observe no drops in 289 calls is $\exp(-289 \times 0.024) \approx 0.001$. Probability for both events to happen is their product or 0.0004. Now suppose that the true drop rate is $x = 0.0005$. If this is so than the probability to have no drops in 289 calls is $\exp(-289 \times 0.0005) \approx 0.87$. However the probability to have such drop rate from lognormal distribution is about 0.00055. So the combined probability is about 0.0005. We see that our guesses were not to good.

To get the most likely drop rate we need to maximize the product of the probabilities in equations (1) and (3) with regard to $\ln(x)$. This requires a calculation (see Appendix B), but the result is simple. The most probable drop rate, $x^*$, is the solution of this equation:

$$\ln(x^*) + Ns^2 \times x^* = \langle\ln(x)\rangle. \quad (4)$$

Unfortunately, it can't be solved analytically, except for one particular case $N=0$: when no calls have been made than the best guess which can be made of the dropped calls rate is the maximum of the lognormal distribution.

Getting back to Bell South in Athens, numerical solution of equation (4) gives $\ln(x^*) \approx -5.3$, or $x^* \approx 0.005$. With this value of the drop rate the probability density of both events to happen together is about 0.03, about 60 times improvement over 2 preceding figures.

What is the accuracy of this estimate? We found the most probable drop rate. How far away from the maximum should we go that probability become considerably less? One can show (see Appendix B) that the inferred probability distribution of the drop rate for a particular market/carrier combination is given by a lognormal distribution with mean $\ln(x^*)$ and dispersion

$$\boldsymbol{s}^* = \boldsymbol{s} / \sqrt{1 + N \times \boldsymbol{s}^2 \times x^*} \ . \quad (5)$$

For the case we are considering this dispersion is $\boldsymbol{s}^* = 1.07 / \sqrt{1 + 289 \times 1.07^2 \times 0.005} \approx 0.66$.

This means that the probability that the drop rate is within the interval $-5.96 = -5.3 - 0.66 < \ln(x) < -4.64 = -5.3 + 0.66$, or $0.0026 < x < 0.0097$, is 68%.

Now the reader might be concerned that the figure above does not match well with the mentioned few paragraphs above probability density 0.03. This is because the total probability to observe zero drops out of 289 calls based on our drop calls statistics is not too high, about 0.046 (see the Appendix C).

The same technique can be used to find the most probable drop rate when it was not measured 0. Maximizing the product of equations 1 and 2 we obtain:

$$\ln(x^*) + (x^* - n/N)N\boldsymbol{s}^2 = \langle \ln(x) \rangle \ . \quad (6)$$

The dispersion is again given by the equation 5, with only difference that now $x^*$ is a solution of the equation (6).

The same analysis applies to blocked calls rate and to no service percentage. As one can see from Figures 4 and 5 and from Table 2 the distributions of failed call rates are close to lognormal. The only difference from the case of dropped calls is different parameters of the lognormal distribution (they are given in the Table 2 in Appendix A)

If the reader remembers the present discussion started from the statement that zero drop rate is impossible because nothing is perfect. This analysis is imperfect as well. It may be that the use of the nationwide statistics is not justified and one would better use market statistics, or may be nationwide statistics, but for particular carrier. However in that case we will end up with poorer statistics to find the parameters of the distribution.

## APPENDIX A

Table 2. Parameters of distributions

| | | Dropped | Blocked | No service |
|---|---|---|---|---|
| Mean | Linear | 0.038 | 0.033 | 0.062 |
| | Logarithm | -3.74 | -3.97 | -3.85 |
| Standard Deviation | Linear | 0.051 | 0.049 | 0.127 |
| | Logarithm | 1.068 | 1.305 | 1.762 |
| Skewness | Linear | 3.2 | 4.0 | 3.6 |
| | Logarithm | .0027 | -0.47 | -0.10 |
| Kurtosis | Linear | 13 | 24 | 15 |
| | Logarithm | -0.34 | 0.071 | -0.63 |

Mean, standard deviation, skewness and kurtosis of the failed calls rates distributions in linear and logarithmic coordinates. Note that we used such offset for kurtosis that for normal distribution it is 0 (not 3 as is sometimes defined [2].)

## APPENDIX B

The inferred probability distribution of the dropped call rate for a particular market/carrier combination where out of $N$ calls $n$ were dropped can be obtained using Bayes equation [3]:

$$p_i(n, N, \ln(x)) = \frac{p(\ln(x)) \times p(n, x, N)}{\int p(\ln(x)) \times p(n, x, N) d \ln(x)} \quad (7)$$

The numerator can be rewritten as:

$$p(\ln(x)) \times p(n, x, N) =$$

$$\frac{1}{\sqrt{2\boldsymbol{p}}\boldsymbol{s}} \times \frac{N^n}{n!} \times \exp(F(N, n, \langle \ln(x) \rangle, \boldsymbol{s}, \ln(x))) \quad (8)$$

where

$$F(N, n, \langle \ln(x) \rangle, \boldsymbol{s}, \ln(x)) =$$

$$-N \exp(\ln(x)) + n \ln(x) - (\ln(x) - \langle \ln(x) \rangle)^2 / 2\boldsymbol{s}^2 \quad (9)$$

Obviously, the probability in Eq. 7 is at maximum when the function $F$ of Eq.9 is at maximum. By differentiating it with regard to $\ln(x)$ and setting the result to zero we obtain Eq.6 (or Eq.4 when $n = 0$).

To obtain an approximation for $p_i(\ln(x))$ we can expand function $F$ of Eq.9 near maximum (see Ref. [4])

$$F(\ln(x)) \approx$$

$$F(\ln(x^*)) + \frac{1}{2} F''(\ln(x^*)) \times (\ln(x) - \ln(x^*))^2 =$$

$$F(\ln(x^*)) - \frac{1 + Nx^* \boldsymbol{s}^2}{2\boldsymbol{s}^2} \times (\ln(x) - \ln(x^*))^2$$

to get

$$p_i(\ln(x)) \approx$$

$$\frac{1}{\sqrt{2\boldsymbol{p}}\boldsymbol{s}^*} \exp(-(\ln(x) - \ln(x^*))^2 / 2(\boldsymbol{s}^*)^2) \quad (10)$$

with $\boldsymbol{s}^* = \boldsymbol{s} / \sqrt{1 + N \times \boldsymbol{s}^2 \times x^*}$, which is Eq.5. The Eq.10 is asymptotically exact when $\boldsymbol{s}^*$ is small and gives a decent approximation when it is less than 1.

## APPENDIX C

The probability to observe no drops in $N$ calls can be obtained by integrating the product of the probabilities in equations 1 and 3 numerically. An alternative is to use the so-called saddle point asymptotic integration [4], result of which is:

$$p(N, 0) \approx \exp(-0.5 \times N \times x^* ((\boldsymbol{s} / \boldsymbol{s}^*)^2 + 1)) \frac{\boldsymbol{s}^*}{\boldsymbol{s}} \ .$$

In our case of no drops in 289 calls the probability is about 4.6%.

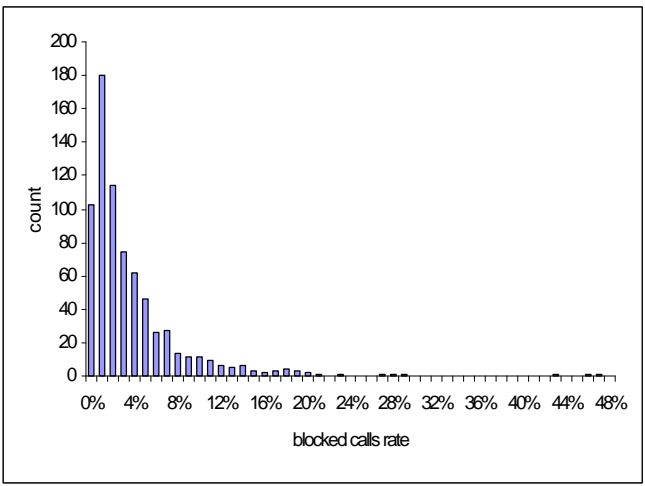
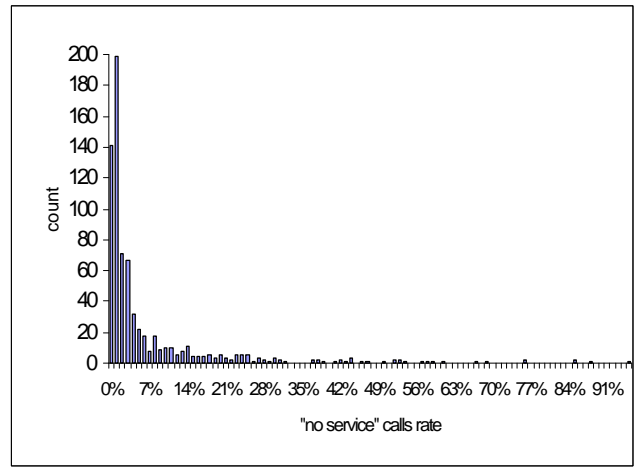

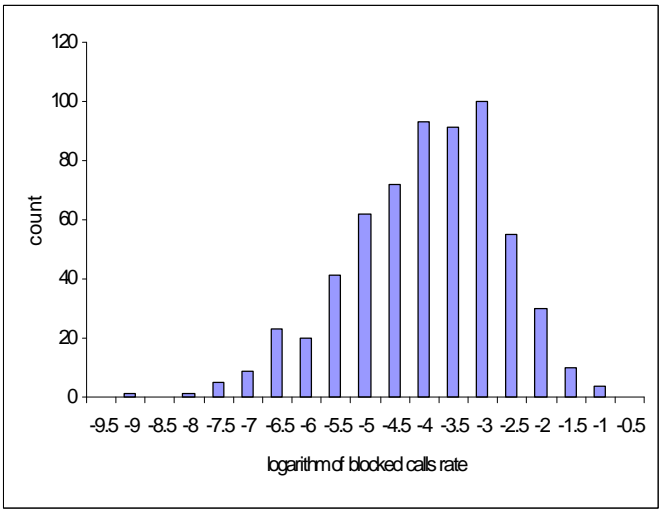
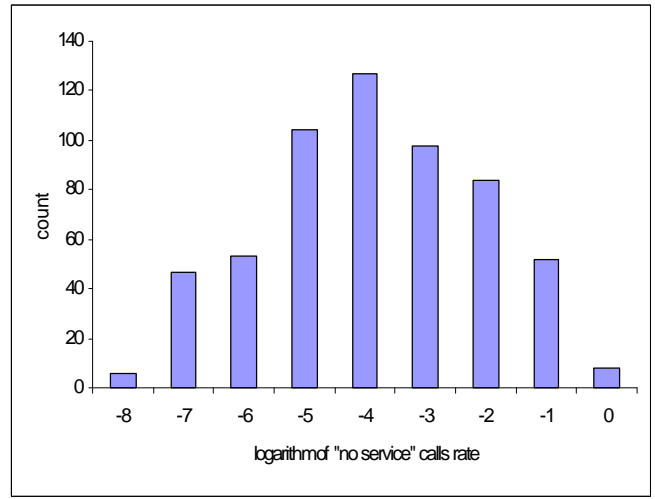

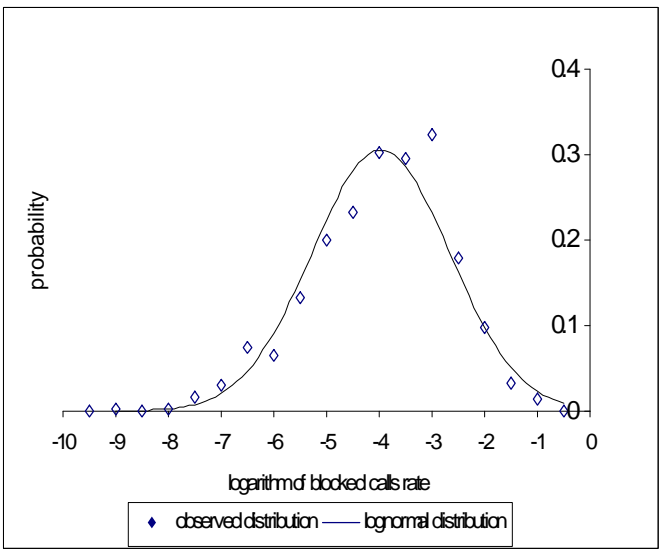

**Figure 4.** Distribution of blocked calls rates.

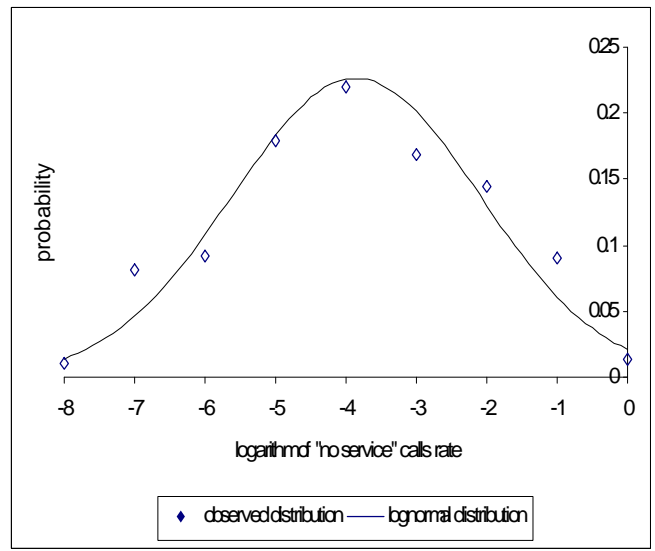

**Figure 5. "**No service" calls rates distributions
.